\documentclass{ws-procs9x6}

\begin{document}

\title{Form Factors of Hadronic Systems in Various Forms  
of Relativistic Quantum Mechanics}

\author{B. DESPLANQUES}

\address{Laboratoire de Physique Subatomique et de Cosmologie \\
(UMR CNRS/IN2P3--UJF--INPG),  \\
 F-38026 Grenoble Cedex, France}

\maketitle

\abstracts{
The form factor of hadronic systems in various forms of relativistic 
quantum mechanics is considered. Motivated by the agreement of the nucleon 
``point-form'' results with experiment, results for a toy model 
corresponding to the simplest Feynman diagram are first presented. 
These ones include the results for this diagram, which plays the role 
of an experiment, for the front-form and instant-form in standard
kinematics ($q^+=0$ and Breit frame), but also in  unconventional 
kinematics and finally a Dirac's point-form inspired approach. 
Results for an earlier ``point-form'' approach are reminded. Results 
are also presented for the pion charge form factor. Conclusions 
as for the efficiency of various approaches are given. }

\section{Introduction}
The primary goal for studying baryon physics is to get insight on how QCD 
is realized in the non-perturbative regime. This includes for instance the 
structure of baryons in terms of constituent quarks and the properties 
of these ones. In this respect, form factors represent  an important source 
of information since their momentum dependence allows one to probe baryons 
at different scales. To fully exploit the experimental data however, a safe 
implementation of relativity is required. 

There are various ways to implement relativity. Ultimately, they should 
converge to unique predictions by incorporating two- or many-body currents 
beside the one-body current generally retained in calculations. 
In the frame of relativistic quantum mechanics, different forms have been 
proposed, following the work by Dirac \cite{Dirac:1949cp}. They can be 
classified according to the symmetries of the hyperfurface which physics 
is described on, determining at the same time the dynamical or kinematical 
character of the Poincar\'e-group generators \cite{Keister:sb}. 
For applications to baryons, some approaches do well 
by giving the constituent quarks some form factor \cite{Cardarelli:1995dc}. 
Other ones do without \cite{Merten:2002nz,Wagenbrunn:2000es}. 
This obviously calls for an independent check of the reliability 
of the underlying formalisms.  

A system that can provide a useful testing ground consists of two scalar 
particles of mass $m$ exchanging a scalar particle of mass $\mu$. 
In the two extremes, $\mu=0$ (Wick-Cutkosky model) and $\mu=\infty$ 
(vertex function), the Bethe-Salpeter equation 
can be solved and solutions can be used to calculate form factors that are 
exact ones and can thus play the role of an ``experiment''. It has also been 
shown that the mass spectrum provided by the first model is reasonably 
described by a simple mass operator \cite{Amghar:2000pc} whose solutions 
can be employed for the calculation of form factors in different forms of 
relativistic quantum mechanics. In the second case, the uncertainty due 
to the range of the interaction is reduced to a minimum and the solution 
is essentially known. The comparison of the results obtained in both ways
(Bethe-Salpeter equation and mass operator) offers many advantages. 
The dynamics of the interaction is the simplest one that can be imagined 
and the uncertainty arising from that one on the effective interaction 
entering the mass operator is limited. Spin effects are absent, allowing 
one to check possibly large effects like those due to the 
Lorentz contraction \cite{Amghar:2002jx} or the ``point-form'' spectator 
approximation at high $Q^2$~\cite{Allen:2000ge}. Finally, the intrinsic 
form factors of the constituents, if any, cancel in the comparison 
while they have to be accounted for when an experiment is involved, 
which could actually contribute to their determination. 

In this paper, we concentrate on the above schematic model in the case 
$\mu=\infty$. Beside the ``experiment'', we consider form factors in various 
forms and, for some of them, with different kinematics. The forms of  
interest here include the instant and front ones as well as a Dirac's 
inspired point form~\cite{Desplanques:2004}. An earlier ``point-form'' 
implementation  \cite{Sokolov:1985jv,Lev:1993,Klink:1998}, 
which differs from the Dirac's one by the fact it involves a hyperplane 
perpendicular to the velocity of the system \cite{Sokolov:1985jv} 
(and not a hyperboloid), is also considered. Some results similar 
to the above ones will be considered for the pion charge form factor. 
Apart from the fact that the pion represents a 
physical system, there are real data but, as already mentioned, one has 
then to worry about constituent form factors.

The plan of the paper is as follows. After reminding the relation 
of the constituent momenta to the total momentum of the system 
in different forms, we present and discuss results  for the charge 
form factor in the schematic model. The emphasis is put on the differences and 
the similarities beween various approaches. This is followed by a presentation 
of results for the pion charge form factor. 
Their discussion and a conclusion are finally given.

\section{Form factors in a schematic model}
\begin{figure}[htb]
\centerline{\epsfxsize=8cm\epsfbox{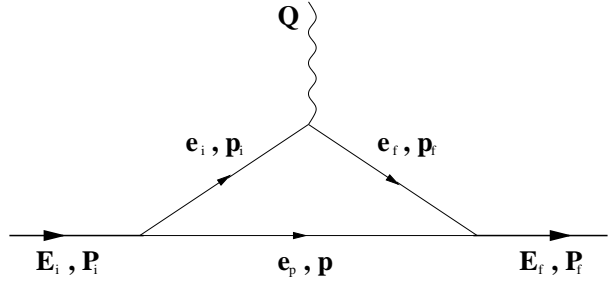}}   
\caption{Kinematics relative to the photon absorption on a two-body system.
\label{fig1} }
\end{figure}  

The interaction of a two-body system with an external 
probe is represented in Fig. \ref{fig1} in the single-particle current 
approximation. In the frame of relativistic quantum mechanics, which we are 
working in here, the particles in the intermediate state are on mass shell 
($e=\sqrt{m^2+p^2}$). In order to calculate the corresponding form factor, 
two ingredients are needed:\\ 
1) the relation of the constituent momenta to 
the total momentum  of the system (respectively $\vec{p}_1$ and $\vec{p}_2$,  
and $\vec{P}$). This one characterizes the form under consideration 
and is closely related to the symmetries of the hypersurface which physics 
is described on. \\
2) a solution of the mass operator which can be taken as form-independent.

In all cases we consider, the relation of the constituent momenta to 
the total momentum can be cast into the form:
\begin{equation}
\vec{p}_1+\vec{p}_2-\vec{P}=(\vec{\xi}/\xi^0) \,(e_1+e_2-E_P),
\label{boost0}
\end{equation}
where $\xi^{\mu}$ is specific of each approach. This relation is fulfilled 
by a Lorentz-type transformation that allows one to express the constituent 
momenta
in terms of an internal variable, 
$\vec{k}$, which enters the mass operator, and the total momentum, $\vec{P}$. 
This transformation, which underlies the Bakamjian-Thomas construction 
of the Poincar\'e algebra in the instant form \cite{Bakamjian:1953kh}, 
reads:
\begin{equation}
\vec{p}_{1,2} = \pm\,\vec{k} \pm\vec{w}\;\frac{\vec{w} \! \cdot \!
\vec{k} }{w^0 +1} + \vec{w}  \;e_k \, ,
 \;\;\;e_{1,2} = w^0\,e_k\pm \vec{w}\! \cdot \!
 \vec{k}  \, ,
\label{boost1}
\end{equation}
with $\vec{w}$ and $w^0=\sqrt{1+\vec{w}\,^2}$ given by: 
\begin{equation}
 w^{\mu} =\frac{ P^{\mu} }{ 2\,e_k }
+ \frac{ \xi^{\mu} }{ 2\,e_k } \; \frac{4\,e_k^2-M^2}{
 \sqrt{ (\xi \cdot P)^2 + (4\,e_k^2-M^2)\;\xi^2 } +  \xi \cdot P}\, .
\label{boost2} 
\end{equation}
The 4-vector $\xi^{\mu}$ appearing in the above expression multiplies 
a term $(4\,e_k^2-M^2)$ that can be transformed into an interaction one. 
This is in accordance with the expectation that changing the 
surface pertinent to each approach implies the dynamics. Apart from these 
features, it can be seen that the above expression is independent of the scale 
of the 4-vector $\xi^{\mu}$. Thus, up to an irrelevant scale,  the 4-vector 
$\xi^{\mu}$ is given as follows: \vspace{1mm}\\
- {\it instant form}: \hspace{0.2cm} 
$\xi^0=1, \;\;\;\;\vec{\xi}=0  ,$ \vspace{1mm}\\
- {\it front form}:  \hspace{0.5cm} 
$\xi^0=1, \;\;\;  \;\vec{\xi}=\vec{n},$  
($|\vec{n}|=1 $, fixed direction),\vspace{1mm}\\
- {\it Dirac's point form}: \hspace{0.1cm} 
$\xi^0=1, \;\;\;\vec{\xi}=\vec{u},  $ ($|\vec{u}|=1$, 
from $(p_1+p_2-P)^2=0$). \vspace{1mm}\\
In the last case, $\vec{u}$ can point to any direction, consistently 
with the absence of any orientation on a hyperboloid \cite{Desplanques:2004}. 
The boost transformation introduced in an earlier ``point-form'' approach 
\cite{Klink:1998} is recovered from  Eq. (\ref{boost0}) by taking 
$\xi^{\mu} \propto P^{\mu}$. Thus, the calculation of a form factor 
in this last approach implies initial and final states that are described 
on different hyperplanes, contrary to all other cases where a unique 
hypersurface is involved.

\begin{figure}[htb]
\centerline{\epsfxsize=5.5cm\epsfbox{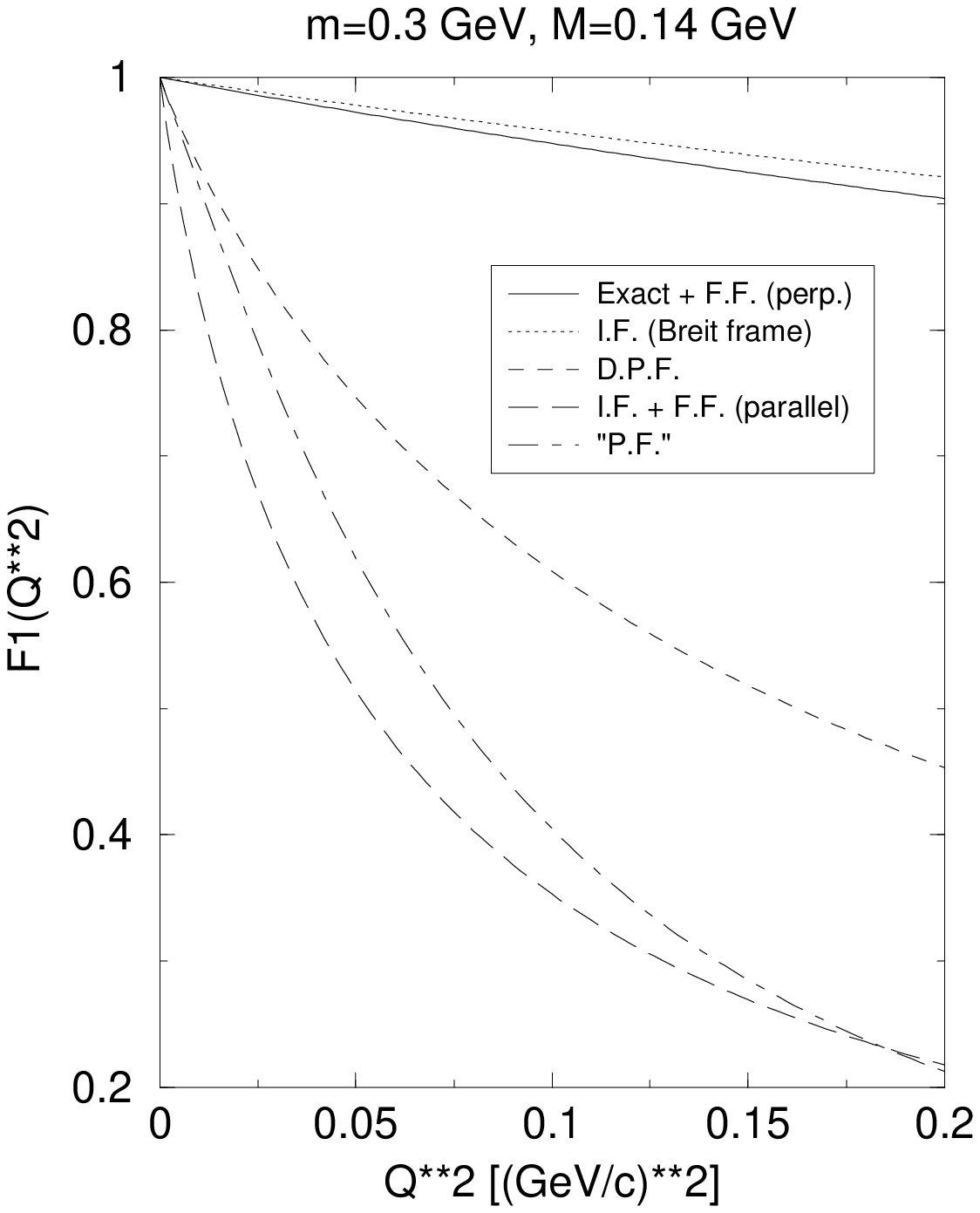} \hspace*{2mm}
\epsfxsize=5.58cm\epsfbox{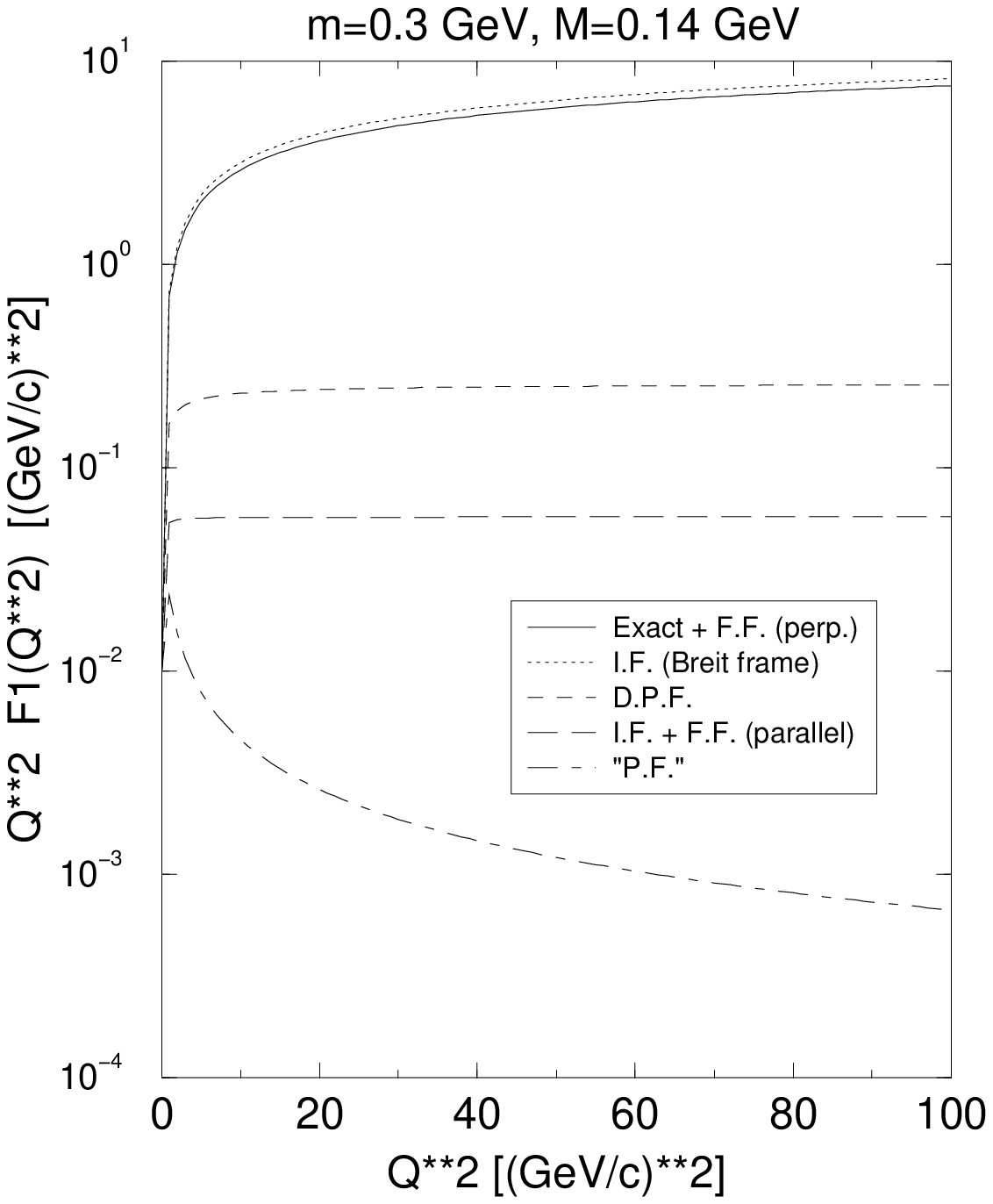}}   
\caption{Charge form factor in various forms of relativistic quantum 
mechanics: left for low $Q^2$ and right for high $Q^2$ (multiplied 
by $Q^2$ in the last case). 
\label{fig2} }
\end{figure}  

The second ingredient needed for the calculation of form factors is 
the solution of a mass operator. For the interaction model considered here 
(exchange of an infinitely-massive boson), the solution can be taken as 
$\phi(\vec{k})\propto (\sqrt{e_k}\; (4\,e_k^2-M^2))^{-1}$ \cite{Amghar:2002jx}. 
The constituent and total masses entering this expression, $m=0.3$ GeV and 
$M=0.14$ GeV, are chosen in accordance with those used  
for the pion results presented in the following section.

Two form factors can be considered for the system under consideration, a charge 
one, $F_1(Q^2)$, and a Lorentz-scalar one, $F_0(Q^2)$. Their expressions for 
different forms, which can be cast into a unique one in terms of the 4-vector, 
$\xi^{\mu}$, can be found in most cases in Ref. \refcite{Amghar:2002jx}. 
Due to the lack of space, results are presented here for $F_1(Q^2)$. 
Two aspects of charge form factors are of interest, the charge radius 
and the asymptotic behavior, which are determined by the low and high $Q^2$ 
parts respectively. The form factors are presented accordingly 
in the left and right parts of Fig. \ref{fig2}. They contain:\\
- the exact form factor (continuous line), \\
- the standard front-form one ($q^+=0$), identical to the 
exact result, \\
- the instant-form one (dotted line, I.F. ( Breit frame)), \\
- a Dirac's inspired point-form one (short-dashed line, D.P.F.), 
correspon\-ding to a fully Lorentz-covariant result \cite{Desplanques:2004}, \\
- a front-form one in the configuration where the initial and final momenta, 
$\vec{P}_i$ and $\vec{P}_f$ are parallel to the front orientation 
(dashed line, F.F. (parallel)),\\
- an instant-form one in the parallel kinematics, 
$\vec{P}_i \parallel \vec{P}_f$, with an average momentum going to $\infty$ 
(I.F. (parallel)) (coincides with the previous curve),\\
- and an earlier ``point-form'' one (dash-dot line, ``P.F.''). \\
As form factors scale in most cases like $Q^{-2}$ (up to log 
terms), the quantities  displayed in the right part of Fig. \ref{fig2} 
are multiplied by the factor $Q^2$. 

Form factors clearly fall into two sets: close or even identical 
to the ``experiment'' and  far apart. The difference in the 
behavior can be ascribed to the dependence on the total mass of the system, $M$ 
\cite{Desplanques:2001ze}.
Rather weak in the former case, it becomes important in the latter one. 
Actually, results in this last case depend on the momentum transfer $Q$ through 
the combination $Q/2M$. This produces a charge radius that scales like the 
inverse of the mass of the system, hence the steep slope of the corresponding 
form factors at small $Q^2$. This feature is also responsible for the 
suppression of the form factors at high $Q^2$ (a factor $\simeq (M/2m)^2$).  
In the  ``P.F.'' case, further suppression occurs,  the dependence 
on  $Q^2$  involving an extra factor $ (1+Q^2/4M^2)$ at high $Q^2$
\cite{Allen:2000ge},  hence the approximate asymptotics $Q^{-4}$ 
of the corresponding form factor in the present case.

Results very similar qualitatively to the above ones are obtained 
for the Lorentz-scalar form factors.  The same conclusion holds 
to a lesser degree for interaction models involving the exchange 
of a zero-mass boson (Wick-Cutkosky model) \cite{Amghar:2002jx}. 
In this case, some uncertainty affects the determination 
of the effective interaction entering the mass operator. With the 
simplest possible interaction, the discrepancy 
does not however exceed a factor 2 in the cases where an identity 
was previously obtained. The discrepancy between the two sets 
of results mentioned above is considerably increased, in relation 
with a different asymptotic behavior, $Q^{-4}$ instead of $Q^{-2}$.

\section{Pion charge form factor}
\begin{figure}[htb]
\centerline{\epsfxsize=5.3cm\epsfbox{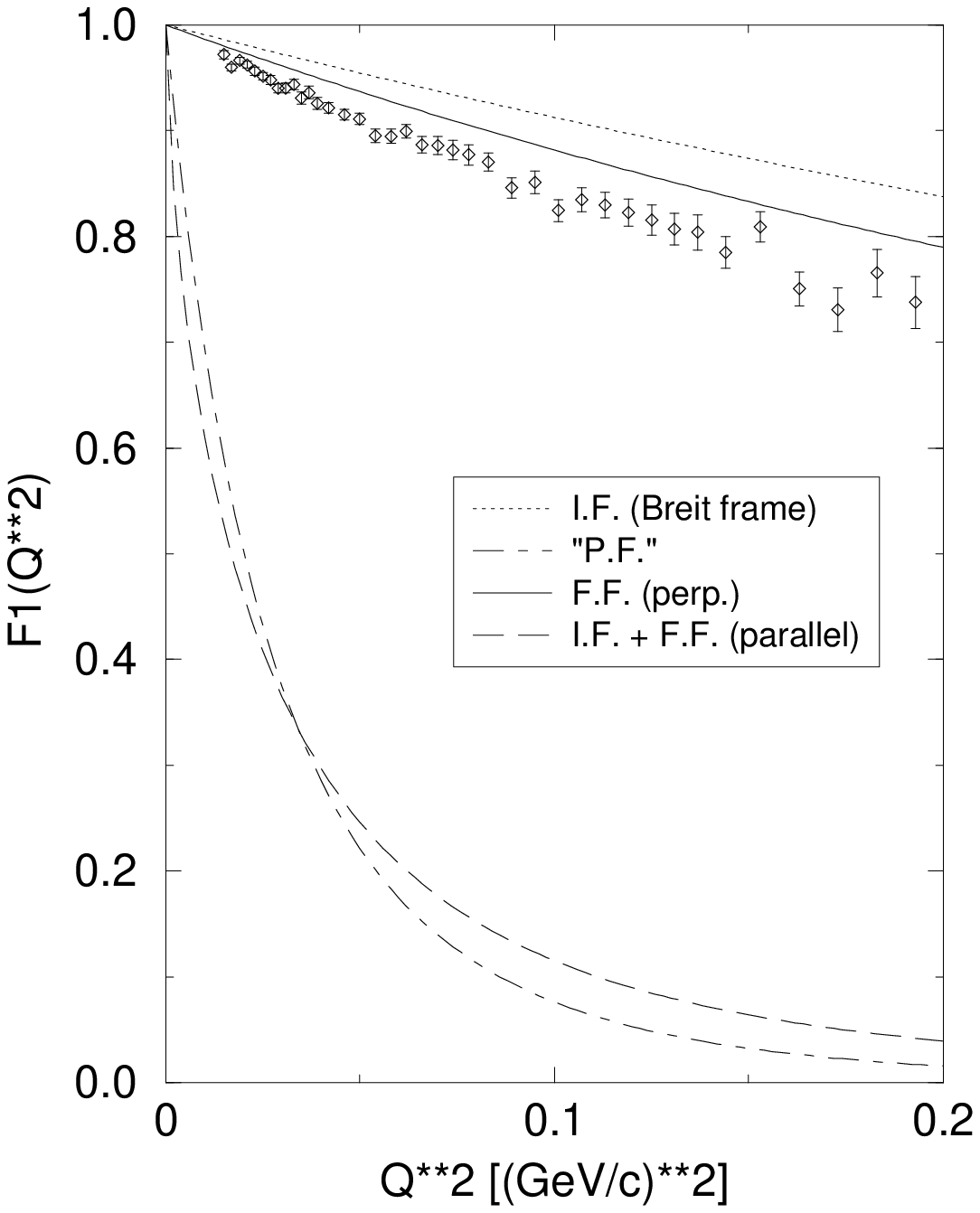} \hspace*{2mm}
\epsfxsize=5.6cm\epsfbox{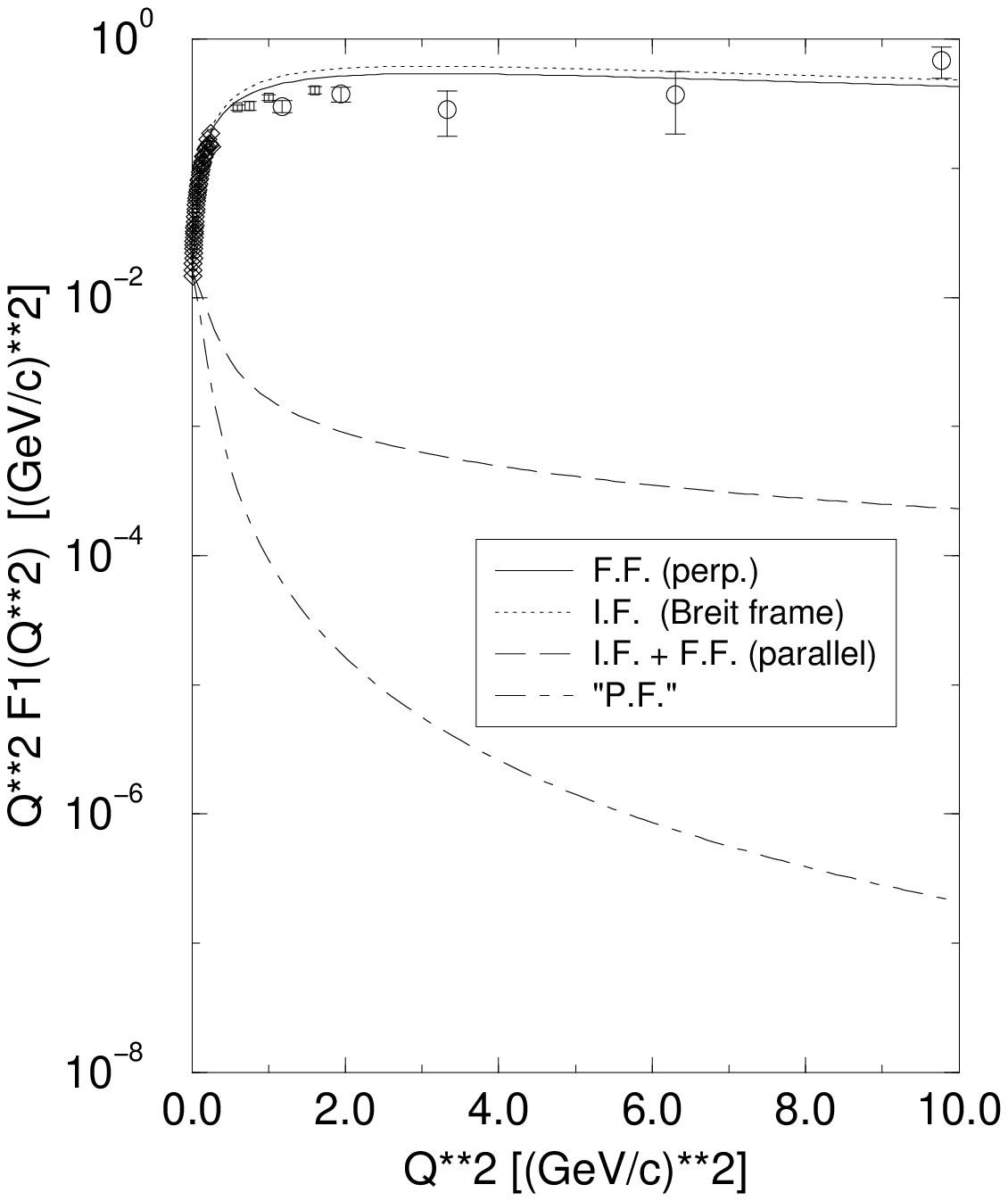}}   
\caption{Same as Fig. \ref{fig2} but for the pion charge form factor. 
\label{fig3} }
\end{figure}  
In this section, we consider the pion charge form factor for which 
experimental data are known. Calculations similar to those of the previous 
section are presented. There are however two main differences that make 
worthwhile to consider this system. The pion consists of two quarks 
that carry a 1/2-spin. The interaction has a finite range and is dominated 
by a one-gluon exchange at small distances. Moreover, there are  
predictions concerning both the charge radius and the asymptotic behavior 
that should be ultimately recovered \cite{Bernard:1988bx,Farrar:1979aw}. 
Expressions of form factors in the single-particle current approximation,  
which have to account for the quark spin,  can be obtained from 
Ref. \refcite{Amghar:2003tx}. As for the interaction entering the mass 
operator, it is taken as the sum of a confining potential with a standard 
string tension (1 GeV/fm) and a gluon exchange with strength 
$\alpha_s=0.35$. No attempt is made in the present work to optimize 
the results which are presented in Fig. \ref{fig3}. Their examination 
shows they are very similar to the scalar-particle ones, 
confirming those obtained in works with a different scope 
\cite{Simula:2002vm,Bakker:2000pk,deMelo:2002yq}. We however notice that, 
in the best case, the $Q^{-2}$ QCD asymptotic behavior is not recovered 
and that relatively standard two-body currents are needed in this order. 
As the discrepancy with experiment evidenced by the other approaches 
we considered reaches orders of magnitude, one can safely discard them 
as efficient ways to implement relativity.

\section{Discussion and conclusion}
In this paper, we compared different forms of relativistic quantum mecha\-nics 
to calculate form factors. We first considered a schematic interaction model. 
The very good or complete agreement with an exact calculation in some cases, 
the disagreement in other ones leaves no doubt about which approach is 
appropriate or inappropriate to get the bulk contribution of form factors. 
A similar conclusion can be inferred from considering the pion form factor. 
Thus, front- and instant-form approaches with standard kinema\-tics 
($q^+=0$ and Breit frame respectively) appear as quite convenient to 
calculate the dominant contribution to form factors. The same approaches 
with unconventional kinematics or the point-form approach require a large 
contribution from non-standard two-body currents (of the type $0/0$). 
 
One may wonder why approaches based on a single-particle current work well 
while other ones, fully covariant in some cases, don't. Taking into account 
that front- and instant-form approaches with unconventional kinematics 
or a Dirac's point-form inspired approach give relatively similar results 
(same asymptotic behavior and dependence on $Q$ through the factor $Q/2M$), 
we are tempted to consider that the approaches that work are an exception. 
A simple argument explaining the above observation would be helpful. 
One often has advantage to break some symmetry to get closer to the 
properties of a physical system (think to a deformed mean field for 
calculating the binding energy of a nucleus with $J=0$). Thus, among the 
different approaches considered here, the Lorentz-covariant one (point form) 
may not be necessarily the best one. This approach has been advocated because 
the corresponding kinematical character of boosts  makes easy to get wave 
functions of states with momenta different from the rest-frame one. 
However, one also has to relate the transferred momentum to the momenta 
carried by the struck particle. In the point-form approach, this relation 
involves the dynamics, while it has a kinematical character in the 
field-theory models underlying physics of interest here. With this 
respect, using the point-form approach represents a bad strategy as 
interaction effects introduced in the above relation will have to be 
removed later on, under the form of two-body contributions. Instead, 
the standard front- and instant-form approaches are those which  better 
fulfill the kinematical character of the above momentum relation in field 
theory. This is perhaps the reason why they are more appropriate for 
the calculation of form factors. 

We began this presentation by reminding that the goal for studying 
form factors is to learn about hadronic physics. Present results for 
the pion charge form factor are still prematurate. However, taking 
into account that only the standard front- and instant-form approaches 
provide a reliable implementation of relativity when a single-particle 
current is considered, it is found that the corresponding form factors 
can {\it a priori} accommodate a reasonable constituent form factor. 
Its precise nature has to be determined. \vspace{4mm}

\section*{Acknowledgments}
We are very grateful to A. Amghar,  S. Noguera and L. Theu{\ss}l for 
their collaboration at the early stage of the development of this work.


\end{document}